# Density Matrix of the Fermionic Harmonic Oscillator

**Batool A. Abu Saleh**
[1]Department of Applied Science, Faculty of Ma'an College, Al-Balqa' Applied University, Al-Salt, Jordan

## Abstract

The path integral technique is used to derive a possible expression for the density operator of the fermionic harmonic oscillator. In terms of the Grassmann variables, the fermionic density operator can be written as: $\rho_F(\beta) = c^*(\beta)c(\beta) \pm c^*(\beta)c(\beta)e^{-\beta\omega}$, where $+(-)$ means that the sum over all antiperiodic (periodic) orbits. Our density operator is then used to obtain the usual fermionic partition function which describes the fermionic oscillator in thermal equilibrium. Also, according to the periodic orbit $c(\beta) = c(0)$, the graded fermionic partition function is obtained.

*Keywords:* Fermionic Density Operator, Path Integral Representation, Fermionic Harmonic Oscillator, Fermionic Partition Function.

## Introduction

Path integrals are common tools used to describe the time evolution of the quantum mechanical system [7,810]. In nature the particles can be classified into bosons and fermions, instead of the commutative variables that describe bosons the classical description of the fermions require the anti-commutative variables called the **Grassmann variables**. Accordingly, there are two kinds of the path integrals; the bosonic path integral and the fermionic path integral. In the Euclidean path integral, the time have no real components, i.e., $t = i\beta$, that means the time evolution operator turns into the quantum statistical operator (density operator) which is a powerful tools for the study of systems in thermal equilibrium $e^{-\beta H}$. Formally; the density operator, whose trace is the partition function $Z(\beta) = Tr[e^{-\beta H}]$ describes the evolution in the imagenary time β. However, in the case of bosonic density operator, some formula have been obtained and interpreted [4,11,1]. In contrast, in the case of fermionic density operator, Siew-Ann in [14] obtain the many-body density matrix $\rho$ in terms of the block Green function matrix, also using the concept of supper matrix Robin [13] obtain the fermionic density supper matrix for the Nuclear Magnetic Resonance (NMR) system. In this paper, we construct the path integral representation for the fermionic density operator at thermal equilibrium $e^{-\beta H_F}$ where $H_F$ being the Hamiltonian of the fermionic oscillator and $\beta = 1/k_B T$, then we obtain an explicit expression for the fermionic density operator which is constructed from the Grassmann variables. Finally, by taking the trace for the expression of the fermionic density operator and with the help of Grassmann algebra, we obtain the usual partition function for the fermionic harmonic oscillator.

Corresponding author: Batool A. Abu Saleh Email: batool.abusalih@bau.edu.jo & batool.abusalih@yahoo.com

### Fermionic Harmonic Oscillator.

Fermions obey the Pauli exclusion principle which states that no two fermions with identical quantum numbers can occupy the same quantum states. Here, we consider a quantum system with a spin - *1/2* particle, this system described by the Pauli matrices $\sigma_x$, $\sigma_y$ and $\sigma_z$ [7]. The fermionic creation and annihilation operators may be written as

$$c^\dagger = \sigma_+ = \begin{pmatrix} 0 & 1 \\ 0 & 0 \end{pmatrix}, \qquad \text{and} \qquad c = \sigma_- = \begin{pmatrix} 0 & 0 \\ 1 & 0 \end{pmatrix}, \tag{1}$$

where $\sigma_\pm = (\sigma_x \pm i\sigma_y)/2$, and the operators $c$ and $c^\dagger$ satisfy the usual anti commutation relations:

$$\{c, c^\dagger\} = cc^\dagger + c^\dagger c = 1, \qquad \text{and} \qquad \{c, c\} = \{c^\dagger, c^\dagger\} = 0. \tag{2}$$

The Fock space (Hilbert space generated by the fermionic creation and annihilation operators) contain only two states $|\emptyset>$ and $|1>$ due to $(c)^2 = (c^\dagger)^2 = 0$. The Hamiltonian which describe the fermionic harmonic oscillator is an operator acting in this space, and it can be written as [7],

$$H = \omega c^\dagger c = \omega N \tag{3}$$

where $N = c^\dagger c$ is the number operator counting the particles (fermions) in the state vector, $N|1> = 1$ and $N|0> = 0$. Accordingly; the eigenvalues of the Hamiltonian operator are given by

$$H|\emptyset> = 0, \qquad and \qquad H|1> = 1 \tag{4}$$

We next present the unnormalized fermionic density operator $\rho_F(\beta)$ for a statistical ensemble which is then defined by

$$\rho_F(\beta) = \sum_{n=0}^{1} |n> e^{-\beta E_n} <n| = |0><0| + |1><1|e^{-\beta\omega} = \begin{pmatrix} e^{-\beta\omega} & 0 \\ 0 & 1 \end{pmatrix} \tag{5}$$

From the above equation, one can easily obtain the partition function of the fermionic oscillator as:

$$Z_F(\beta) = Tr[\rho_F(\beta)] = 1 + e^{-\beta\omega} \tag{6}$$

Incidentally, if one takes the supper trace of the fermionic density operator, one obtains

$$Str[\rho_F(\beta)] = Tr[(-1)^N \rho_F(\beta)] = Tr\left[\begin{pmatrix} -1 & 0 \\ 0 & 1 \end{pmatrix}\begin{pmatrix} e^{-\beta\omega} & 0 \\ 0 & 1 \end{pmatrix}\right] = 1 - e^{-\beta\omega} \tag{7}$$

It is clear that the above expression is related to the inverted bosonic partition function, namely, the graded fermionic partition function [15].

### Grassmann Variables.

As we know fermionic oscillator obeys canonical anti commutation relations, so do a path integral calculation of a fermionic oscillator require the classical anti commuting variables. In this section, we introduce a short review about this variables. The Grassmann variables ($c_i$) satisfy the anticommutation relations

$$\{c_i, c_j\} = c_i c_j + c_j c_i = 0 \tag{8}$$

so that $c_i c_i = 0$ and hence the differential operator $\partial/\partial c_i$ acts on a function as same as the ordinary differential operator

$$\frac{\partial}{\partial c_i} c_j = \delta_{ij} \qquad (9)$$

Integration with respect to a Grassmann variable $c$ is defined according to Berezin [2] to be identical with differentiation

$$\int dc = 0, \quad \rightarrow \quad \int dc\, c = \mathbf{1}, \quad \rightarrow \int dc\, f(c) = \frac{\partial f(c)}{\partial c}, \qquad (10)$$

The Gaussian integral of $n$ Grassmann variables is defined as

$$\int \prod_{j=1}^{n} dc_j^* dc_j\, \exp\left(-\sum_{ij} c_i^* M_{ij} c_j\right) = \det \mathbf{M} \qquad (11)$$

where $\mathbf{M}$ is a real anti-symmetric matrix. Also, the identity operator and the trace were defined in terms of Grassmann variables as

$$1 = \int dc^*\, dc\, |c><c| e^{-c^*c} \qquad (12)$$

$$\text{Tr}[\mathcal{A}] = \int dc^*\, dc\, e^{-c^*c} <-c|\mathcal{A}|c> \qquad (13)$$

So far we defined the most tools that we need. As for the reader who is mostly interested in the Grassmann variables, see for example [4,16].

## Path Integral Formulation of the Fermionic Density Matrix

Here, we want to discuss the path integral formulation of the density matrix for a system described by anti-commuting variables. Consider a fermionic oscillator, the matrix elements of the quantum statistical (time evolution) operator can be written as

$$\mathcal{K}(c,\beta;c',0) = <c|e^{-\beta H_F}|c'> \qquad (14)$$

Note the Euclidean (imaginary) time $\beta$ is used rather than the Minkowski time $t$ by performs a Wick rotation $t \rightarrow i\tau$. However, one can divides the total propagation imaginary time $\beta$ into $N$ elementary steps of interval $\beta/N$ and then using $(N-1)$ times of the identity operator Eq.(12) in terms of coherent states to obtain,

$$\mathcal{K}(c,\beta;c',0) = \int \left(\prod_{k=1}^{N-1} dc_k^* dc_k\, e^{-c_k^* c_k}\right) \prod_{k=1}^{N} <c_k|e^{-\frac{\beta}{N}H_F}|c_{k-1}> \qquad (15)$$

where we have defined $c_0 = c'$ and $c_k = c$. As N→∞, each matrix element in the above equation gives

$$<c_k\left|e^{-\frac{\beta}{N}H_F}\right|c_{k-1}> = e^{c_k^* c_{k-1}} e^{-\frac{\beta\omega}{N} c_k^* c_{k-1}} \qquad (16)$$

In our derivations, we used the fermionic Hamiltonian Eq.(3). Substituting Eq.(16) into Eq.(15), one gets

$$\mathcal{K}(c,\beta;c',0) = \int (\prod_{k=1}^{N-1} dc_k^* dc_k \, e^{-c_k^* c_k}) \exp\{-\sum_{k=1}^{N} -c_k^* c_{k-1} + \frac{\beta\omega}{N} c_k^* c_{k-1}\}$$

$$= \lim_{N\to\infty} \int \left(\prod_{k=1}^{N-1} dc_k^* dc_k\right) \exp\{-\frac{\beta}{N}\sum_{k=1}^{N}\left[\frac{c_k^*(c_k - c_{k-1})}{\frac{\beta}{N}} + \omega c_k^* c_{k-1}\right] + c_N^* c_N^*\}$$

$$= \int \mathcal{D}[c^*,c] \exp\left\{\left(-\int_0^\beta d\tau \, c^*(\partial_\tau + \omega)c\right) + c^*(\beta)c(\beta)\right\} \tag{17}$$

where $\mathcal{D}[c^*,c] = \prod_{k=1}^{N-1} dc_k^* dc_k$. Noting that the fields $c^*(\tau)$ and $c(\tau)$ entering into the path integral are antiperiodic. The partition function for the fermionic oscillator may be obtained by taking the trace of Eq.(17) with the anti-periodic boundary conditions $c(0) = -c(\beta)$ and $c^*(0) = -c^*(\beta)$ and using the generalized zeta function regularization [6]. However, to obtain the density matrix for the fermionic oscillator, we will use the path integral formalism to find the possible solution of the imaginary time propagator in Eq.(17), the action term in this equation can be written as:

$$S[c^*(\tau),c(\tau)] = c^*(\beta)c(\beta) - \int_0^\beta d\tau \, \ell(c^*,\dot{c}^*,c,\dot{c}) \tag{18}$$

where $\ell(c^*,\dot{c}^*,c,\dot{c})$ is the classical Euclidean Lagrangian of the fermionic oscillator, which is defined as [5],

$$\ell(c^*,\dot{c}^*,c,\dot{c}) = c^*\dot{c} + \omega c^* c \tag{19}$$

Substituting Eq.(19) into Eq.(18), we obtain:

$$S[c^*(\tau),c(\tau)] = c^*(\beta)c(\beta) - \int_0^\beta d\tau \, (c^*\dot{c} + \omega c^* c) \tag{20}$$

one can easily simplified the second term in this action by using the following change of variables as:

$$\eta(\tau) = c(\tau)e^{\omega\tau} \quad \text{and} \quad \eta^*(\tau) = c^*(\tau)e^{-\omega\tau} \tag{21}$$

The Jacobian of these change of variables equal one (Jean, 2013). Accordingly, Eq.(20) may be written as:

$$S[\eta^*(\tau),\eta(\tau)] = \eta^*(\beta)\eta(\beta) - \int_0^\beta d\tau \, \eta^*(\tau)\dot{\eta}(\tau) \tag{22}$$

The time integral can be replaced by a discrete form by dividing the imaginary time interval $[0, \beta]$ into $N$-sub-intervals of duration $\Delta\tau = \beta/N$ such that

$$\int_0^\beta d\tau \, \eta^*(\tau)\dot{\eta}(\tau) \approx \sum_{n=0}^{N-1} \eta_{n+1}^* \frac{\eta_{n+1} - \eta_n}{\Delta\tau} \Delta\tau \tag{23}$$

By expanding the summation, except for the end-points $n = 0$ and $n = N$, we find

$$\sum_{n=0}^{N-1} \eta_{n+1}^*\{\eta_{n+1} - \eta_n\} = \cdots - \eta_{m+1}^* \eta_m + \eta_m^* \eta_m - \eta_m^* \eta_{m-1} + \cdots - \eta_2^* \eta_1 \tag{24}$$

Using the Grassmann algebra and Berezin integration [2] to integrate the exponential of these terms over $\eta_m$ and $\eta_m^*$, we obtain

$$\mathcal{K}(\eta_{m+1}, \eta_m) \to \int\int d\eta_m^* d\eta_m \exp\{\eta_{m+1}^*\eta_m - \eta_m^*\eta_m + \eta_m^*\eta_{m-1}\}$$

$$= \int\int d\eta_m^* d\eta_m \{(1 + \eta_{m+1}^*\eta_m)(1 - \eta_m^*\eta_m)(1 + \eta_m^*\eta_{m-1})\}$$

$$= \int\int d\eta_m^* d\eta_m \{\eta_m\eta_m^* + \eta_m\eta_m^*\eta_{m+1}^*\eta_{m-1}\}$$

$$= 1 + \eta_{m+1}^*\eta_{m-1} = \exp\{\eta_{m+1}^*\eta_{m-1}\} \tag{25}$$

The $N$-$1$ integrations over the pair $\eta_n \eta_n^*$ for each time point in the summation (24) except the end-points, gives the propagator of the fermionic oscillator as:

$$\mathcal{K}(\eta, \beta; \eta, 0) = \exp\{\eta^*(\beta)\eta(\beta) - \eta^*(\beta)\eta(0)\} \tag{26}$$

Here, as $N \to \infty$, we replace the subscript $N$ by continuous imaginary time label ($\beta$). By using Eqs.(21) the propagator of the fermionic oscillator can be expressed in terms of $c$'s as:

$$\mathcal{K}(c, \beta; c, 0) = \exp\{c^*(\beta)c(\beta) - c^*(\beta)c(0) e^{-\beta\omega}\} \tag{27}$$

this propagator represents the evolution of the fermionic oscillator through the imaginary time interval [0, $\beta$]. Now we want to interpret Eq.(27) as a statistical mechanical density matrix. Incidentally, in the bosonic case, see for example [4] the bosonic (graded bosonic) partition function is obtained by taking the trace of the bosonic density matrix by integrating over all states such that $x_i = x_f$ ($x_i = -x_f$). In our case the fermionic propagator is constructed from the classical anti-commuting variables (Grassmann variables), so with the help of the properties of Grassmann variables, the fermionic density matrix (27) can be written in the following form

$$\mathcal{K}_\pm(c, \beta; \pm c, \beta) = \{1 + c^*(\beta)c(\beta) \mp c^*(\beta)c(\beta) e^{-\beta\omega}\} \tag{28}$$

where $\mathcal{K}_+(c, \beta; c, \beta)$ and $\mathcal{K}_-(c, \beta; -c, \beta)$ represent the sum over all the periodic and anti-periodic orbits respectively. The fermionic density matrix (28) with the anti-periodic orbit $c(0) = -c(\beta)$ is related to the fermionic partition function while with the periodic orbit $c(0) = c(\beta)$ it may be related to the graded fermionic partition function which is obtained by imposing the periodic boundary condition on the measure of the fermionic path integral [5]. However, we will now evaluate the partition function of the fermionic harmonic oscillator from Eq.(28) for both periodic and antiperiodic paths.

$$Z_F^{(\pm)}(\beta) = \int dc(\beta)dc^*(\beta)\{1 + c^*(\beta)c(\beta)\{1 \mp e^{-\beta\omega}\}\}$$

$$= 1 \mp e^{-\beta\omega} \tag{29}$$

Therefore

$$Z_F^{(-)} = 1 + e^{-\beta\omega} = \sum_{n=0}^{1} e^{-\beta\omega n} = \sum_{n=0}^{1} <n|e^{-\beta H_F}|n> \tag{30}$$

$$Z_F^{(+)} = (1 - e^{-\beta\omega}) = (\sum_{n=0}^{\infty} e^{-\beta\omega n})^{-1} = (\sum_{n=0}^{\infty} <n|e^{-\beta H_B}|n>)^{-1} \tag{31}$$

where $Z_F^{(\pm)}(\beta)$ is the partition function corresponding to the boundary conditions $c(\beta) = \pm c(0)$. These results is in exact agreement with one obtained in [8,9]. As a consequence these results emphasize that the inverted partition function for periodic boundary condition equal to the partition function of the bosonic harmonic oscillator which has energy eigenvalues for the n*th* mod given by $E_n = n\omega$, where $n = 1, 2, 3, \ldots$. Furthermore, from the fermionic partition function, one can obtain all other observables, for instance the entropy, the free energy, the average energy and so on.

## Conclusion

In this paper, we used the path integral technique to find an explicit solution of the fermionic path integral which represent the quantum statistical propagator (fermionic density matrix) of the fermionic harmonic oscillator. The obtained expression for the density matrix represents one state problem, i.e., $(2 \times 2)$ matrix, thus it may be generalized to an arbitrary number of available states. The method used in this study was to construct the fermionic density operator for a very simple system (fermionic oscillator) that may be solved explicitly but was rich enough to apply this method into more complicated systems. Our expression of the fermionic density operator is then used to obtain the well-known fermionic (graded fermionic) partition functions of the Fermi gas by integrating over the Grassmann variables with the antiperiodic (periodic) orbits.